\documentclass[12pt]{article}
\usepackage{geometry}                % See geometry.pdf to learn the layout options. There are lots.
\usepackage{graphicx}
\usepackage{amssymb}
\usepackage{epstopdf}
\usepackage{color}
\usepackage{wrapfig}
\usepackage{lipsum}

\def\bea{\begin{eqnarray}}
\def\eea{\end{eqnarray}} 
\def\be{\begin{equation}}
\def\ee{\end{equation}}
\newcommand{\ba}{\begin{eqnarray}}
\newcommand{\ea}{\end{eqnarray}}

\def\tbh{T_{bh} } 
\def\abh{A_{bh} } 

\def\tds{T_{cos} } 
\def\tcos{T_{cos} } 
 
\def\acos{A_{ds} } 

\def\rbh{r_{bh} }
\def\rds{r_{cos} }
\def\rcos{r_{cos} }
\def\qt{ \tilde{a} }
\def\at{ \tilde{a} }

\def\bk{\hfill\break}

\begin{document}

\begin{titlepage}
\vfill
\begin{flushright}
ACFI-T15-14
\end{flushright}

%\vfill
\vskip 3.0cm
\begin{center}
\baselineskip=16pt
{\Large\bf Cosmography of KNdS Black Holes }
\vskip 0.1in
{\Large\bf and Isentropic Phase Transitions }

\vskip 0.3cm

{\bf James McInerney ${}^{a}$, Gautam Satishchandran${}^{a,b}$, Jennie Traschen${}^{a}$}

\vskip 0.2cm
${}^a$ Amherst Center for Fundamental Interactions, Department of Physics\\ University of Massachusetts, Amherst, MA 01003\\

\vskip 0.1in ${}^b$ Department of Physics, University of Chicago, Chicago, Il 60637\\ 
\vskip 0.1 in Email: \texttt{jpmciner@umass.edu, gautamsatish@uchicago.edu, traschen@physics.umass.edu} 
\vspace{6pt}
\end{center}
\vskip 0.2in
\par
\begin{center}
{\bf Abstract}
 \end{center}
\begin{quote}
 We present a new analysis of Kerr-Newman-deSitter black holes in terms of thermodynamic quantities that are defined in the observable portion of the universe;
between the black hole and cosmological horizons.
In particular, we replace the mass $m$ with a new `area product' parameter $X$.
The physical region of parameter space is found analytically and thermodynamic quantities
 are given by simple
 algebraic functions of these parameters.
 We find that different geometrical properties of the black holes are usefully distinguished by  the sum of the black hole and cosmological entropies.
The physical parameter space breaks into a region in which the total entropy, together with $\Lambda$, $a$ and $q$
uniquely specifies the black hole, and a region in which there is a two-fold degeneracy. In this latter region, there are isentropic pairs of black holes, having the
 same $\Lambda$, $a$, and $q$, but different $X$. The
  thermodynamic volumes and masses differ in such that there  are high and low density branches.
 The partner spacetimes are related by a simple inversion  of $X$, which has a fixed point at the state of maximal total entropy.
 We compute the compressibility at fixed total entropy and find that it diverges at the maximal entropy point. Hence a picture emerges of high and low density
 phases merging at this critical point.

 \vfill
% \hrule width 5.cm
\vskip 2.mm
\end{quote}
\end{titlepage}

\section{Introduction}

The elegant Kerr-Newman spacetimes, describing asymptotically flat,
rotating, charged black holes, have been extensively studied in contexts ranging from astrophysics through quantum gravity. The extension of Kerr-Newman to cosmological constant $\Lambda>0$ \cite{Carter:1973}
has also been the subject of extensive study, for example, in the contexts of  black hole thermodynamics 
\cite{Gibbons:1976ue,Gibbons:1977mu,Gibbons:1979xm,Davies:1987ti,Davies:1989ey,Guven:1990,Romans:1991nq,Ghezelbash:2004af,Sekiwa:2006qj, Anninos:2010gh}, 
black hole pair production via instantons  \cite{Mellor:1989wc,Mann:1995vb,Booth:1998gf,Booth:1998pb,Dias:2003st,Dias:2004rz}, 
 quasi-normal modes  \cite{Mellor:1989ac,Yoshida:2010zzb}, and dS/CFT \cite{Strominger:2001pn,Cai:2001sn,Cai:2001tv,Ghezelbash:2001vs,
 Bousso:2001mw,
 Jing:2002aq,Balasubramanian:2002zh,Guijosa:2003ze,Setare:2004hw,Guijosa:2005qi,Chen:2010bh,Chen:2010jj,Xiao:2014uea,Anous:2014lia,Chatterjee:2015pha,Sarkar:2014dma,Sarkar:2014jia}.   
 DeSitter black holes have  also been proposed as nucleation sites 
for the decay of a metastable vacuum state \cite{Gregory:2013hja,Burda:2015isa,Burda:2015yfa},  
while primordial black holes produced during inflation may have played a role in early universe physics and left observable signatures \cite{Page:1976wx,Carr:1976zz,
Barrow:1991dn,Carr:1993aq}\footnote{On this topic we have given a few references covering a range of issues. The interested reader can consult
the hundreds of citations of these foundational works.}. 
More generally, the Kerr-Newman-deSitter (KNdS) metric remains one of the few analytic tools available for studying black holes in a cosmological setting.
While the KNdS metric closely resembles  the asymptotically flat case, the 
 non-zero cosmological constant dramatically changes the causal structure of the spacetime, and consequently its mechanics and thermodynamics.
Analyses  of the geometric properties of the KNdS spacetimes are given in \cite{Romans:1991nq,Brill:1993tw,Akcay:2010vt}, while the structure of asymptotically
dS spacetimes is studied in \cite{Ashtekar:2014zfa} and the extension of the KNdS metric
 to general dimensions is given in \cite{Gibbons:2004uw}. The work we present here contributes to mapping out the rich structure of the 
 KNdS metrics with an emphasis on the role played by the combined entropy of the black hole and deSitter horizons.
 
Black holes with $\Lambda >0$ pose several theoretical challenges that are distinct from $\Lambda \leq 0$. DeSitter black holes have both a cosmological horizon, with associated temperature $\tcos$ and entropy $S_{cos}$,
and a black hole horizon, with temperature $\tbh$ and entropy $S_{bh}$.  In general these two temperatures are distinct, conflicting with the usual
condition for thermodynamic equilibrium.  
 The conserved mass  $m$ of a deSitter black hole is generated by a Killing field in the usual way.  However,
 $m$ is defined at future infinity, rather than spatial infinity.   An
observer therefore has to `wait a long time' to measure the mass, rather than `getting far away'. 
Moreover,  the relevant Killing vector is spacelike at future infinity, rather than timelike.
Hence $m$ does not so obviously
 correspond to the thermodynamic energy of the system. 
 
 In response to these issues,  the thermodynamics of stationary black holes with $\Lambda>0$ was formulated
in \cite{Dolan:2013ft} in terms of geometrical properties of the region between the black hole and cosmological horizons. 
In that work, the mass term in
  the first law and Smarr relation is replaced by a contribution from the thermodynamic volume $V$ between the horizons,
  times its conjugate variable $\Lambda$ (see the discussion and equations (\ref{smarr}) and (\ref{first}) below). 
The two temperatures and entropies enter in a symmetrical way, as do the contributions from the charge and angular momentum. Motivated by these results,
we go on to study the properties of KNdS black holes in terms of parameters likewise defined in the observable portion of the universe.
We retain the cosmological constant $\Lambda$,  the rotational parameter $a$, and the charge $q$, but we replace the mass $m$ by an `area product' parameter $X$. This allows us to make considerable progress in analyzing the cosmography KNdS black holes\footnote{We use the word `cosmography' as shorthand for the geometric and thermodynamic properties of the spacetime.}.
For Schwarchild-deSitter (SdS) black holes, the new variable $X$ essentially reduces to the thermodynamic volume $V$. In the general case with rotation and charge,
$V$ is a simple algebraic function of $X$. 

The total entropy $S_{tot}$,  one-quarter the sum of the black hole and cosmological horizon areas, 
 turns out to be a useful  tool for organizing the cosmography of  KNdS spacetimes. It is considerably more
 difficult to identify the  restrictions on the parameters in the KNdS metric, compared to the KN metric, such that the spacetime actually
 describes a black hole. Indeed, this has involved numerical work to invert the non-linear relations between the natural algebraic
 parameters and the physical ones \cite{Booth:1998gf}.  We will show that by
 replacing $m$ with  $ X$ the physical parameter space can be identified analytically. The allowed parameter space scales with $l$, the
deSitter curvature scale defined by $\Lambda =3/l^2 $, and
  has the structure ${\cal P}\times R_p$,
where ${\cal P}$ contains the values of points $p=(a/l , q/l)$ and $R_p$ is a finite length segment of the allowed values of $X/ l^2$ at fixed $p$.  
A finite range of values of $S_{tot}$ are generated at each $p$ as
 $X$ varies along $R_p$.  Considering these values of the entropy, we show that in one region of ${\cal P}$, the black hole is uniquely fixed
 by specifying $S_{tot}$. In the remainder of ${\cal P}$ a degeneracy occurs. 
 Smaller values of  $S_{tot}$ correspond to a unique black hole, but for larger values of $S_{tot}$  there are two distinct black holes with the same entropy. One 
 black hole has a smaller thermodynamic volume $V$, while the other has larger $V$. The entropy reaches a maximum value when the pair of configurations
 coincide. Interestingly, the degenerate states of total entropy are related to each other by the simple map 
 \begin{equation}\label{inversion}
 X\rightarrow ( l^2\sqrt{a^2 + q^2} ) /X.
 \end{equation}
  The maximal entropy configuration occurs at the fixed point of this map. 
 We compute the compressibility of KNdS black holes and
 find that it diverges at the maximal entropy state.  The picture then is of two black hole phases, composed of isentropic pairs,  which merge at a critical point.
 
Another advantage of trading the mass $m$ for the area product parameter $X$ is that 
   the system is cast into a more tractable form. With previous parameterizations it is difficult to answer questions such as  how the two horizon
 temperatures and  areas vary with $m$. Here
   we are able derive
    simple algebraic  formulas for the quantities $\tbh$, $\tcos$, $\abh$, $\acos$,  $V$, and $m$ in terms
   $\Lambda ,\  a, \ q$ and $X$. Representative plots are given below. We also give
    embedding diagrams for degenerate black hole  geometries, {\it i.e.}  geometries which have the same boundary area but different
    volumes, as well as a linked video showing a sequence of the horizons of KNdS black holes with fixed $\Lambda,\  a,\ q$ as $X$ varies
    across its allowed range. 
The symmetry  of the entropy  under inversion (\ref{inversion}) significantly simplifies the analysis.
  We have not yet found a fundamental underlying principle for this symmetry. It may be that the  dS/CFT 
 can shed light on this issue.

 This paper is organized as follows. In Section (\ref{basics}) we review the KNdS metric, introduce the new parameter $X$, and derive
 algebraic formulas for thermodynamic quantities in terms of $X$. In Section (\ref{SdSbhs}) we specialize to the 
case of Schwarzschild-deSitter black holes with $a=q=0$, which turns out to be particularly simple in terms of the new variable $X$. 
In Section (\ref{kndsbhs}) we derive the physical parameter space and
analyze the entropy and other thermodynamic properties for the general KNdS black holes. In Section (\ref{pictures}) embedding
diagrams are presented, and in
section (\ref{compress}) we compute the compressibility and discuss the isentropic phase transition. Section (\ref{end}) contains concluding discussion.

\section{KNdS black holes and the area product parameter}\label{basics}
In this section we briefly review the first law and Smarr relations for stationary, charged black holes with $\Lambda > 0$, 
when the thermodynamics is formulated in the region between the black hole and cosmological horizons. Then we record needed
details of the KNdS metric and proceed to
introduce the area product parameter $X$. Lastly, algebraic expressions for the horizon  areas, temperatures, volume, and mass,   in terms
of $\Lambda ,\  a,\  q$ and $X$ are derived. 

\subsection{General thermodynamic relations}
This work was motivated by the successful use of  the thermodynamic volume in studies of AdS black holes, and investigating
whether this would also be the case for dS black holes. The volume $V$ was found to arise naturally in the Smarr relation and first law for
AdS black holes as the variable 
conjugate to $\Lambda$  \cite{Kastor:2009wy}, and was then used in in studying
isoperimetric inequalities \cite{Cvetic:2010jb}. Further research made use of $V$ in deriving various equations of state, and 
analyzing phase transitions for AdS black holes 
\cite{Dolan:2010ha,Dolan:2011xt,Kubiznak:2012wp,Altamirano:2013ane,Altamirano:2013uqa,Altamirano:2014tva,Dolan:2014jva}.
As in the earlier analysis of the Hawking-Page phase transition \cite{Hawking:1982dh} and
subsequent work which explored the extended AdS-black hole phase space \cite{Caldarelli:1999xj,Chamblin:1999tk,Chamblin:1999hg},
the temperature plays a key role.
For example, studies of AdS black holes typically proceed by looking at the 
  isotherms of the black holes in the pressure -  volume phase space,
  where the pressure is provided by the cosmological constant
  \begin{equation}\label{lambda}
{2\Lambda \over (D-1) (D-2)} ={1\over l^2 } \  , \quad P= -{\Lambda \over 8\pi }
\end{equation}
 Critical temperatures and associated phase transitions can then be identified by seeing 
 whether, or not, an isotherm develops extrema as a function of the temperature.
Since AdS itself does not have a naturally specified temperature, the formalism of classical thermodynamic ensembles carries over to Euclidean AdS,
where the Euclidean time period is fixed by the black hole temperature. The free energy can then be used to quantify properties of the
phase transition. 

For black holes in deSitter none of these tools carries over in a simple way. There are two temperatures, $\tcos$ and $\tbh$,  associated with the 
cosmological and black hole horizons respectively, so it is not clear how to treat the system as a fluid with one temperature, or what temperature
to use for a free energy. The Smarr formula and the first law (equations (\ref{smarr}) and (\ref{first}) below)
 treat each temperature and each horizon entropy on an equal footing. There are special cases when the
 two temperatures are equal, and in these cases, calculations of the Euclidean action give that it is equal to $(\acos +\abh)/4$, that is, to
 the total horizon entropy \cite{Gibbons:1976ue,Gibbons:1979xm,Hawking:1982dh,Mellor:1989wc,Booth:1998pb,Booth:1998gf}. 
 Very interestingly, recent calculations done for cases when the temperatures are not equal give the same
  result \cite{Gregory:2013hja,Burda:2015isa}.
These considerations lead us to use the entropy as the quantity to organize the properties of the black holes, rather than temperature,
an approach which turns out to be quite useful.

When  $\Lambda > 0$ a Smarr formula for stationary charged  black holes can be derived which only involves quantities on the black hole
and cosmological horizons \cite{Dolan:2013ft}, which in $D=4$ is 
\begin{equation}\label{smarr}
 T_{bh} S_{bh} +   T_{cos}S_{cos} 
   = {1\over 8\pi} \Lambda V - ( \Omega _{bh}  - \Omega_{cos} ) J - {1\over 2} \Phi Q
\end{equation}
where $J$ is the angular momentum, $Q$ is the charge, $S_h= A_h /4$, $\Omega_h$, and $T_h$ are the horizon
entropies, rotational velocities, and temperatures respectively, and  $\Phi = \phi_{bh}- \phi_{cos}$ is  the electric potential difference between the two horizons.
 In our notation the temperatures are always positive with $2\pi T_h = |\kappa _h |$. The thermodynamic
volume between the horizons $V$ is equal to $V_{cos}-V_{bh}$, the difference between the thermodynamic volumes associated with each of the
cosmological and the  black hole horizons separately.  
The mass does not appear since it subtracts out from two distinct  Smarr relations which connect the black hole (cosmological) horizon to infinity.
Similarly, the first law formulated in the region between the two horizons is \cite{Dolan:2013ft}
\begin{equation}\label{first}
 T_{bh}\delta S_{bh} +    T_{cos}\delta S_{ds}  =
  -  {1\over 8\pi} V \delta \Lambda  -   ( \Omega _{bh}  - \Omega_{cos} ) \delta J  -\Phi \delta Q 
\end{equation}
These fundamental relations, which hold for any stationary Einstein-Maxwell-positive $\Lambda$ black hole spacetime,
 illustrate that there are more relevant thermodynamic quantities for dS than AdS. We now specialize to the known
 analytic KNdS metrics.

\subsection{Kerr-Newmann-deSitter black holes}

In $D=4$ the KNdS metric for a rotating, charged black hole with positive cosmological constant is 
\cite{Carter:1973}
\begin{eqnarray}\label{rotmetric}
ds^2   =  & &  -\frac{\Delta}{\rho^2}\left(dt-\frac{a\sin^2\!\theta}{\gamma} d\varphi\right)^2 +\frac{\rho^2}{\Delta} dr^2
+\frac{\rho^2}{\Psi} d\theta^2 \\ \cr \nonumber
& & +
\left(1+{a^2 \over l^2 } \cos^2\!\theta \right) \frac{\sin^2\!\theta}{\rho^2}\left(adt-\frac{r^2+a^2}{\gamma}d\varphi\right)^2 
\\   \nonumber
\end{eqnarray}
where
\ba\label{delta}
\Delta&=&(r^2+a^2)(1- {r^2\over l^2 } )-2mr+ q^2  \ \\ \cr \nonumber
 \Psi &=&1+{a^2 \over l^2 } \cos^2\!\theta  \,,\\
\rho^2&=&r^2+a^2\cos^2\!\theta\,,\quad  \gamma =1+a^2 / l^2  \\ \nonumber
\ea
The gauge field is $A = -q r/ \rho^2 (dt\!- { a\over \gamma} \sin^2 \theta  d\varphi ) $.
The ADM mass $M$,  angular momentum $J$, and electric charge $Q$  are related to the metric parameters $m, a$ and $q$ by
$M=m / \gamma^2 , \  J= am / \gamma^2 $, and $ Q= q /\gamma$. 
The angular velocities and the electric potential at each horizon is given by 
$  \Omega_{h}= a \gamma/ (r_{h}^2+a^2 ) , \   \phi_{h} = q  r_{h} / (r_{h}^2+a^2 )$.
 Details can be found in \cite{Mellor:1989wc}, and an analysis  of the global spacetime structure is given in \cite{Akcay:2010vt}. 

The horizon areas and the thermodynamic volumes associated with each of the black hole and the cosmological horizons  are given by
\begin{equation}\label{bhdsav}
 A_{h} = {4\pi \over \gamma} (r_{h}^2+a^2)   \  , \quad  V_{h}=\frac{r_h A_h}{3} + {4\pi \over 3} aJ 
\end{equation}
One sees that $V_{cos}$ and $ V_{bh}$  are not equal to their respective geometric volumes \cite{Cvetic:2010jb}, but they differ by the same angular momentum dependent
 term. Hence, amazingly, the thermodynamic volume $V =V_{cos} - V_{bh}$ between the horizons is equal to the geometric volume \cite{Cvetic:2010jb},
\be\label{rotvol}
V = \frac{1}{3}\left(r_{cos}  A_{cos}-r_{bh} A_{bh}\right) 
\ee
The surface gravity at a horizon is given by $2 \kappa_h = (r_h^2 +a ^2) ^{-1} \Delta^\prime (r_h )$. 
The surface gravity is positive at the black hole horizon and negative at the deSitter horizon.
We define the temperatures as the positive quantities $2\pi \tbh =\kappa _{bh}$ and $2\pi \tds = - \kappa _{dS}$.
The horizons  $r_h$ occur at $\Delta (r_h ) = 0$, which implies that at each horizon 
\begin{equation}\label{zeros}
2m = - {r_h ^3 \over l^2 }  + r_h (1 - {a^2 \over l^2 } ) +{\qt^2 \over r_h } 
\end{equation}
Computing the temperatures and using (\ref{zeros}) to eliminate $m$ gives
\be\label{rottemps}
4\pi(r_{h}^2+a^2) T_{h} =
 \pm \left( (1- {a^2 \over l^2} )r_{h}  -3{ r_{h}^3 \over l^2 }  -{\at^2 \over r_{h} } \right) 
\ee 
where the plus sign is for the black hole horizon $r_{bh} $, the minus sign is for $ r_{cos} $, and
\be\label{atdef}
 \at^ 2 = q^2 + a^2
\ee

\subsection{Trading $m$ for $X$}

The KNdS spacetimes are a four parameter family, written above in terms of $(l,a,q,m)$. We next show that the
system can be more readily analyzed if $m$ is traded for a different  geometric parameter $X$, which we are lead to as follows. 
We define
\be\label{xlatdef}
 X=r_{ds}r_{bh}\  , \quad L= r_{ds} - r_{bh} \  , \quad     
\ee
 Subtracting (\ref{zeros})  for each of $r_h= r_{cos} , \  r_{bh}$ gives 
\begin{equation}\label{rdiffs}
  (r_{cos}^3  -r_{bh}^3 ) =  l^2  (\rds -\rbh  ) \left[ 1 - {a^2 \over l^2 }  - {\qt^2 \over r_{cos} r_{bh} } \right]
\end{equation}
This becomes a relation for $V$ in terms of $X$ and $L$ by making use of (\ref{bhdsav}) and (\ref{rotvol}), 
\begin{equation}\label{volagain}
 V = {4\pi \over 3\gamma} l^2 L\left( 1- {\qt ^2\over X} \right)
\end{equation}
The volume will be non-negative in the physical parameter space, which will be derived in  Section (\ref{kndsbhs}). Next, note that 
the quantity $( r_{cos} - r_{bh} )$ can be factored out of equation (\ref{rdiffs}) yielding $ ( r_{cos} - r_{bh} )^2  = l^2 - a^2  -3  r_{cos}  r_{bh} 
  -   \qt^2 /( r_{cos} r_{bh} )$, or
\begin{equation}\label{givesl}
L^2 = l^2 -a^2 -3X - {\qt^2 l^2 \over X}
\end{equation}
 We will view equation (\ref{givesl}) as giving $L$ in terms of $X, q, a, l$, and so $V$ is also given as a function of 
these parameters in equation (\ref{volagain}).
 
Since $ X=r_{bh} ( r_{bh} +L) $ the horizon radii are given in terms of $ X$  by 
\begin{equation}\label{qarad}
r_{bh/cos}   =\mp {L\over 2} + {1\over 2} \sqrt{ L^2 + 4X } 
\end{equation}
where the minus sign is for $\rbh$ and the plus sign for $\rcos$.
 The horizon areas follow from the radii. Of particular interest in this work is the sum of the areas of the black hole and deSitter horizons,
 which has the simple form
\be\label{totalarea}
{A_{tot} \over 4\pi} =  l^2 \gamma^2  - \gamma \left( X + { \qt ^2 l^2 \over X } \right) 
\ee
Note that the total area is invariant under replacing $X$ with $\at ^2 l^2 / X$,
\be\label{lsymm}
A_{tot}  \left( {\at^2 l^2 /  X }\right) = A_{tot}  (X)
\ee
This symmetry will prove to be useful and interesting.

Lastly we find in terms of $l,a,q$, and $X$. Denote the inner black hole or Cauchy horizon as $r_{in}$
and the negative root as $r_{neg}$, so that 
$\Delta (r) $  factors as
\be\label{factordelta}
 \Delta (r) = -{1\over l^2} (r- \rds ) (r-\rbh ) (r-r_{in} ) (r- r_{neg} )
 \ee
Comparing to (\ref{delta}) and ,aking use
of the expressions (\ref{qarad}) for $\rds$ and $\rbh$ gives
\be\label{rotmass}
m=  { \sqrt{L^2 + 4X}  \over 2 l^2 } \left( X +{\qt ^2 l^2 \over X} \right)
\ee
Hence the metric functions are now specified in terms of $l,a,q$ and $X$, so that  $X$ replaces $m$ as the fourth metric parameter.

The main results of this section are the relations (\ref{volagain}) through (\ref{rotmass}) which give thermodynamic quantities
as functions of the physical parameters $l,a,q,$ and $X$. 

The geometrical significance of $X$ is that it is related to the product of the horizon areas by
\begin{equation}\label{xarea}
 X^2 = {\gamma ^2  \over 16\pi ^2 } \left( A_{ds} - {4\pi \over \gamma} a^2 \right) \left( A_{bh} - {4\pi \over \gamma } a^2 \right) 
\end{equation}
However, a reader might legitimately point out that each of $\rcos $ and $\rbh$ has a similar geometric significance  so why should we go through the bother of introducing
their product $X$? Good question!  First, $X$ is only one parameter whereas the two radii would be a redundant set. Suppose one decides
to use $\rbh  $ along with $l, a, q$, and wants to graph the areas, temperatures, volume, and mass as functions of $l,a,q$ and $\rbh$.
The area and temperature of the black hole horizon are given by direct substitution into (\ref{bhdsav} ) and (\ref{rottemps}).
To find the other quantities one needs $\rcos$.
Fixing $\rbh$ determines $m$ through $\Delta (\rbh ) =0$. Once $m$ is known one needs to solve the quartic equation  $\Delta (\rcos) =0$
to find $\rcos$. Therefore in order to plot, for example, the entropies
as a function of $\rbh$, a quartic has to be solved for each value of $\rbh$. If one chooses $m$ as the fourth parameter rather than $\rbh$
the situation is worse - then a quartic needs to be solved to find each of $\rbh$ and $\rcos$. On the other hand, equations
(\ref{givesl}) through (\ref{rotmass}) give the thermodynamic quantities directly in terms of $l,a,q$ and $X$ and so can be plotted directly.
Indeed, the total entropy (\ref{totalarea}) is simple enough to be accurately sketched using basic calculus.

 Once  the physical range of parameters is established (Section(\ref{kndsbhs})) it is straightforward to show that  $\abh $ and $m$ are  
are monotonically related to  $X$, so that qualitatively one can simply think of large $X$ as large $\abh$ and {\it vice-versa}. 
Lastly, we note that the physically relevant range of $X$ will be found by algebra no more difficult than solving a quadratic.
This is in contrast to the situation if  the four horizon roots are used as parameters, in which case the physically allowed  parameter space
must be found numerically \cite{Booth:1998gf}.

The temperature $ \tbh$ ($\tcos$)  is given as a function of $l,a,q$ and $  \rbh$ ($\rcos$)  in (\ref{rottemps}), so with the 
formula for the radii in equation (\ref{qarad}) each temperature is known in terms of the four parameters $l,a,q,$ and $X$. This allows one to 
determine both temperatures for a given metric by simple substitution. Plots of the temperatures in some representative  cases are given in Figure 1.
Of special interest in the literature are the equal temperature 
solutions with $\tcos =\tbh$. This requires  that  $ - (\rbh ^2 + a^2 ) \Delta^\prime (\rds  ) = (\rds ^2 + a^2 ) \Delta^\prime (\rbh  )$.
One finds that solutions occur when either $\rds =\rbh$, which is when the two horizons coincide and the common temperature is zero, or 
$X^2 -2a^2 X -\at^2 l^2 =0$, with positive solution
\be\label{xeq}
X_{eq} = a^2 +\sqrt{ a^4 + \at^2 l^2 }
\ee
 In the analysis of reference \cite{Romans:1991nq}, which studied the charged, non-rotating system, these were referred to as ``lukewarm" black holes.
Not all choices of $(l, a, q)$  give values of $X_{eq}$ that are in the physical parameter space.
 As a result only in part of $(l, a, q)$ parameter space are there
black holes with (non-zero) equal temperatures.
Figure 1 gives plots of $\tbh$ and $\tds$ as a function of $X$ in two representative cases from two different regions of parameter space.
At the small $X$ side the black hole temperature starts at zero, which is the extremal black hole case where the black hole outer and inner horizons
merge.  At the large $X$ end $\tds$ and $\tbh$ go to the common value of
zero as the black hole and cosmological horizons  approach each other. On the left hand side of Figure 1, $\tbh$ is always less than $\tds$, whereas on the right hand side,
there is an equal temperature spacetime, and then $\tbh$ is greater than $\tcos$ for $X> X_{eq}$.

 With $a=0$
the equal temperature metrics occur when 
 $M=Q$. In the general case the mass is given by substituting $X_{eq}$ into (\ref{rotmass}), but does not give such a simple result.
  When $\Lambda =0$ and $a=0$ the condition that the charge is equal to the mass occurs when the inner and outer black hole horizons
 coincide, with common temperature equal to zero. This situation admits multi-centered static black holes described by the Majumdar-Papepetrou metric
 \cite{MP}, illustrating a gravitational-electromagnetic force balance.
 With $\Lambda > 0$ and $a=0$, the charge equal to mass  case  corresponds to the black hole and cosmological temperatures being equal and non-zero.
Further, $Q=M$ still describes a kind of force balance situation -
 the ground state condition was exploited to find multi-centered cosmological $Q=M$ black holes in \cite{Kastor:1992nn}, which is a time dependent generalization of the $Q=M$ Majumbar-Papepetrou  spacetimes. It would be very interesting to find a multi- rotating and charged black hole cosmology, possibly corresponding 
 to the equal temperature situation, but to our knowledge an analytic solution has not yet been found.

\begin{figure*}
\centering
\begin{tabular}{cc}
{\includegraphics[width=0.47\textwidth,height=0.27\textheight]{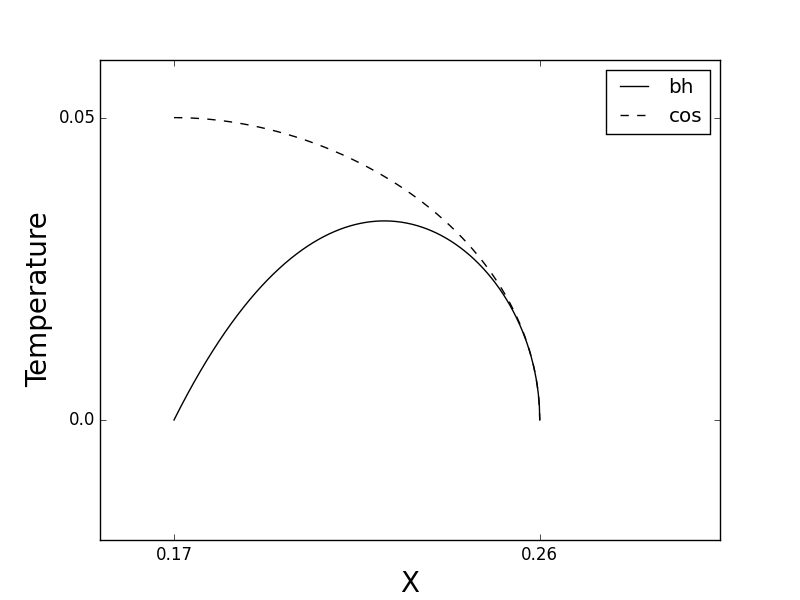}} &
%\rotatebox{-90}{
\includegraphics[width=0.47\textwidth,height=0.27\textheight]{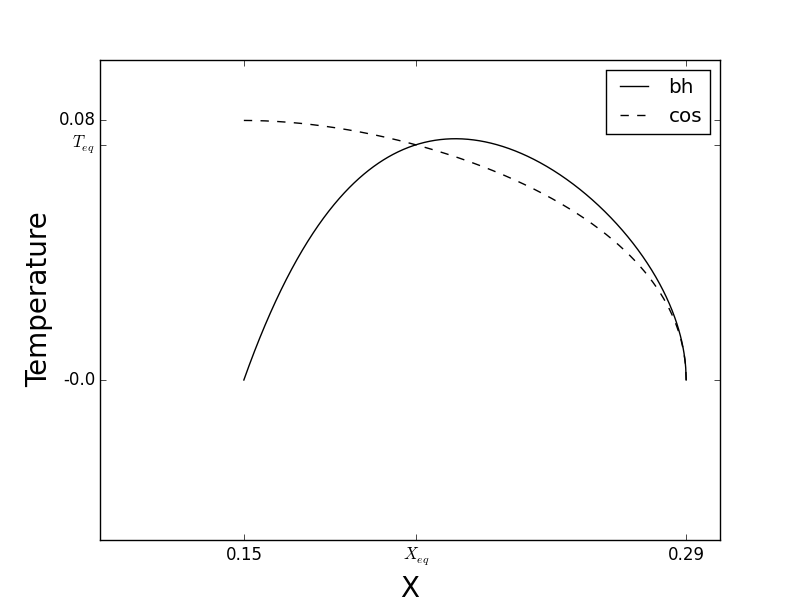}\\
\end{tabular}
\caption{{  $\tbh$ and $\tds$:}
The horizon temperatures are plotted as a function of $X$ for fixed $p= (a/l , q/l )$. On the left $p$ is chosen from 
  the region $\cal{LD}$ in parameter space (see Fig. 3). There are no non-zero equal temperature
        spacetimes. On the right, $p$ is chosen from the domain $\cal{SD}$. In this case 
        $ X_{eq}$ is in the physical parameter space and the curves cross at a non-zero $T_{eq}$. }
\label{fig1}
\end{figure*}

\section{Properties of Schwarzchild-deSitter black holes}\label{SdSbhs}
We first specialize to the Schwarzchild-deSitter black hole metric, in which case the formulae above simplify  considerably. 
Setting $a=0,\  q=0$, the metric (\ref{rotmetric}) reduces to 
\begin{equation}\label{metric}
ds^2 = - f(r) dt^2 +{dr ^2 \over f(r) } +r^2 d\Omega ^2
\end{equation}
where
\begin{equation}\label{fdef}
f(r) =1 -{ r^2 \over l^2 } -{2m\over r } 
\end{equation}
and the thermodynamic volume becomes  
\begin{equation}\label{vdef}
V = {4\pi \over 3}\left( {\rds ^3  -  \rbh ^3 }\right)
\end{equation}
which is the same result as subtracting two Euclidean balls of radii $\rcos$ and $\rbh$ \cite{Dolan:2013ft}. 
With $\at =0$, equation (\ref{volagain}) tells us that
the thermodynamic volume is  proportional to the difference in the two radii. Expressing quantities in units of $l$ we have
\begin{equation}\label{relation}
{1\over l} L  =    {3V\over 4\pi  l^3} \equiv \mu
\end{equation}
Here we have defined the dimensionless volume parameter $\mu$ which takes values $0\leq \mu \leq 1$ to streamline subsequent formulae.
 The expressions for the two horizon radii in (\ref{qarad}) reduce to 
\begin{equation}\label{horrad}
r_{bh/cos} = \mp   {l\mu \over 2}+ {l\over \sqrt{3} } \left[ 1 - {1 \over 4  } \mu^2 \right]^{1/2}\  
\end{equation}
 The deSitter radius takes its maximum value when $\mu =1$, with $\rds =l$ and $\rbh \rightarrow 0$. As
the volume decreases, $\rds$ decreases and $\rbh$ increases to the limiting common value
of $l/ \sqrt{3}$ which is the Nariai limit.

Each horizon area follows from the radii, and the sum of the areas is especially simple,
\begin{equation}\label{horarea}
A_{tot} = 8\pi { l^2 \over 3} \left( 1 + {1 \over 2 }\mu^2  \right)
\end{equation}
which is a monotonic increasing function of $\mu$. At fixed $l$, the total entropy is largest for large volumes, $\mu \rightarrow 1$, which
corresponds to very small black holes.  

The mass as a function of volume is
\begin{equation}\label{mass}
m= {l\over 3\sqrt{3} } \left( 1 - {\mu ^2 \over 4  } \right)^{1/2}  \left( 1- \mu ^2   \right)
\end{equation}
The mass takes its largest  value of $l/(3\sqrt{3}) $  at $\mu =0$ when the horizons approach each other and the volume goes to zero. The mass decreases
monotonically as $\mu$ increases to one, corresponding to very small black holes.
Since both horizons change their area as $\mu$ changes, it is also interesting to see how $m$ varies with $\abh$ alone.
One finds that $\abh$  increases monotonically as $\mu$ decreases, that is,
small volume corresponds to a large black hole horizon. Hence $A_{bh}$ increases with $m$, as is true for
$\Lambda \leq 0$. 

Lastly we look at the temperatures. Taking 
 the sum and difference of  equations (\ref{rottemps}) with $q=a=0$ one finds that the combinations of
  $\rds$ and $\rbh$ which arise are  easily translated to the $l ,\mu$ parameters, giving
  \begin{equation}\label{easytemps}
4\pi (\tds + \tbh ) ={6\mu \over l} 
 \left[ {1-{1\over 2} \mu^2 \over 1-\mu^2  }\right] \  , \quad 4\pi (\tbh - \tds )  ={2\sqrt{3} \over l } \mu^2
 \left[ { (1-{1\over 4} \mu ^2 ) ^{1/2}   \over 1-\mu^2  }\right]
\end{equation}
The individual temperatures follow from (\ref{easytemps}). 
The temperatures are both equal to zero when the horizons coincide as $\mu$ goes to zero (the Nariai limit).
As $\mu$ approaches one, $\tbh$ goes to infinity as for a small Schwarchild 
black hole and $\tds$ goes to its pure deSitter value of $1/(2\pi l)$.
 Plots confirm that  as $\mu$ increases from zero to one, the temperatures
 $\tbh$ and $\tcos$  also increase monotonically. Hence unlike Schwarzchild-AdS
  black holes, for a given value of $\tbh$ (or of $\tcos$) there is only one black hole with that temperature.

\section{Properties of Kerr-Newman deSitter black holes}\label{kndsbhs}
In an anti-deSitter spacetime,  the analysis of the possible black hole states makes use of the tools of
classical thermodynamics. Most famously, the Hawking-Page phase transition  analysis of 
Schwarzchild -AdS black holes starts by looking at black hole isotherms and 
  identifying at what temperatures there are two, one, or no black hole solutions \cite{Hawking:1982dh}.   
  Since the thermodynamics only involves one temperature, $\tbh$, and the system is static, one can work in Euclidean AdS with the Euclidean time period 
   equal to $2\pi / \tbh$,  which then acts like a heat bath for the black hole. Stability analysis proceeds by computing the free energy and 
   specific heat. The AdS / CFT correspondence has prompted a significant amount of further work on phase transitions of charged and rotating AdS black holes,
  as well as on identifying effective equations of state of the black hole system which have analogs in classical thermodynamics.
  Following the introduction of the thermodynamic volume \cite{Kastor:2009wy} studies of  the equation of state and phase transitions of AdS black holes
  have proceeded in a manner similiar from classical thermodynamics, for example by plotting isotherms on $P-V$ diagrams and utilizing the Gibbs
  free energy \cite{Caldarelli:1999xj,Chamblin:1999tk,Chamblin:1999hg,Chen:2010bh,Chen:2010jj,Cvetic:2010jb, Kubiznak:2012wp,Altamirano:2013ane,
Altamirano:2013uqa,Altamirano:2014tva,Dolan:2014jva}.

 There are problems with carrying this program over to black holes in deSitter. When $\Lambda >0$ there
    are two temperatures which are generically not equal, which means that the system
   does not fit into a usual equilibrium thermodynamic description.  It is not  clear if one should consider the
 two temperature relations (\ref{rottemps}) as equations of state for two components of a fluid, or if some linear combination describes an effective
 fluid with an effective temperature \cite{Urano:2009xn,Ma:2013aqa,Zhao:2014zea,Zhao:2014raa,Ma:2014hna,Guo:2015waa},
  or if neither of these approaches captures the critical physics.  In recent work a  combination Gibbs free 
 energy is introduced to study phase transitions \cite{Kubiznak:2015bya} and found features in common with AdS black holes. Our approach is a bit different. Rather than
 temperature, we will use the total entropy $S$ as an organizing concept,
and have found that this is quite useful. 

We briefly summarize what is done in this section.
The analysis starts by identifying the physical  parameter space ${\cal P}$ for the values of $(a/l ,q/l )$.
We then find the allowed values of $X/ l^2$ at each point, which is a finite range. Hence the KNdS parameter space can be expressed in 
units of the curvature radius $l$, and  consists of ${\cal P}$ cross a line segment whose
endpoints depend on the point  in ${\cal P}$. The range of $l$ is $0<l<\infty$, so the full parameter space is given by ${\cal P} \times R_{p} \times \mathbb{R}^+$.
However for brevity we will often suppress the range of $l$.
As $X $ ranges over its allowed values at each $(a/l ,q/l )$ 
 a finite range of values of the total entropy $S$  are obtained.
We show that the set ${\cal P}$ divides into a region in which each value of $S$ corresponds to a unique value of $X$, and
 hence a unique black hole, and another region such that at each $(a/l ,q/l )$ there is a range of values of $S$ that each correspond to 
 two distinct black hole configurations. The two degenerate entropy branches are characterized by corresponding to smaller or larger black holes
 and the same values of $l,q,a$. Equivalently, the isentropic black hole pairs are distinguished by having larger or smaller thermodynamic volume.
  The members of each pair are related to each other by the simple map $X \rightarrow \at ^2 l^2 / X$.
 The branches merge  at the fixed point of this map $X=\at l$, which is the configuration of maximal entropy. 
  
It is worth noting that the algebra below requires nothing more complicated than the quadratic formula.

\subsection{Constraints on parameters}

In this section we find the constraints on the parameters $l, a, q$, and $X$
 so that the metric (\ref{rotmetric}) describes an asymptotically deSitter black hole.
One requires that the spacetime has  cosmological, black hole, and inner Cauchy horizons, and that the coordinate $t$ is 
timelike between the  black hole and cosmological  horizons. 
In the form of the KNdS metric (\ref{rotmetric}) this means that  $\Delta (r) \geq 0$ for $\rbh \leq r \leq \rcos$ and that $T_{bh/cos} \geq 0$.
Such an analysis is carried out in reference \cite{Booth:1998gf} by finding
restrictions on the four roots $r_h$ to $\Delta (r_h )=0$, and then translating the allowed parameter space for the roots to the 
physical parameter space for $l,a,q$, and $m$. This involves a combination of algebraic and numerical work, resulting in the 
elegant plot displayed in Figure 2 of that paper.

Here we will be able to proceed more directly. Consulting equation (\ref{qarad}) we see that the horizon radii are positive real quantities
if $X\geq 0$,  $L^2 \geq 0$, and we define $L$ to be the positive square root of $L^2$. 
We next impose these conditions, plus the condition $T_{bh/cos} \geq 0$,
  and map out the complete parameter space of $a, q,$ and $X$ for a given value of $l$.
  The quantity $L^2$ given in equation (\ref{givesl}) can be factored as  $L^2 = ( X-X_{L-})(X_{L+} -X )/X $ where
\begin{equation}\label{xl} 
 X_{L\pm} = {l^2\over 6} \left(1-{a^2 \over l^2 }  \pm 
\left[ \left(1- {a^2 \over l^2 } \right)^2 - 12  {\at^2 \over l^2  }  \right]^{1/2} \right)
\end{equation} 
Here $X_{L\pm}$ denote the zeros of $L^2$, which are real and positive as long as $1- a^2 / l^2  > 0$ and  $(1- a^2 / l^2  )^2 - 12 (\at / l )^2   \geq 0$. 
The second inequality implies the first. The case of equality
 defines the elliptical-like curve $12{q^2 \over l^2} + 14{a^2 \over l^2 } - {a^4 \over l^4} =1$
 and the allowed points $p =( a/l , q/l)$ in the parameter space ${\cal P}$ lie in the  interior of, or on, this curve. Rewriting this relation in factored form gives
\begin{equation}\label{pset}
{\cal P} = \left\{ ({a\over l}, {q\over l}) :  12 {q^2 \over l^2 }\leq  \left( {a^2 \over l^2 } - (2- \sqrt{3} )^2 \right)  \left( {a^2 \over l^2 } - (2+ \sqrt{3} )^2 \right) 
 \right\}
\end{equation}
 Hence the largest that
 $a/l$ can be is $(2- \sqrt{3} )$. 
We will see in a moment that the additional condition that the temperatures are positive  does not further restrict $l, a, q$, so 
${\cal P}$ is  the allowed parameter space of $(l,a,q)$.
Hence if  $ X_{L-} \leq X \leq X_{L+}$ and $(a,q,l)$ belong to ${\cal P}$ then $X$ and $L^2$ are real and non-negative. This ensures
that the roots satisfy $\rcos \geq \rbh \geq 0$.

 In order that the metric function $\Delta (r)$ be positive between $\rbh$ and $\rcos$, we need to check that 
$\tbh ,\  \tds \geq 0$. Equation (\ref{rottemps}) for the temperatures can be written as
\begin{equation}\label{thfactored}
4\pi{r_{h}(r_{h}^2+a^2) T_{bh/cos}} =
 \mp 3(r_{h}^2-X_{L-})(r_{h}^2-X_{L+}) 
\end{equation} 
were the minus (plus) sign is for the black hole (cosmological) temperature. It is straightforward to check that
assuming $L^2 >0$,
if $\tbh >0$  then $\tcos >0$. Hence it is sufficient to determine the conditions for $\tbh >0$. Assuming that $L^2 >0$, then
 equation (\ref{thfactored}) implies that $\rbh^2 $ must lie in the range $ X_{L-} \leq \rbh^2  \leq X_{L+} $. 
Pursuing some algebra and using (\ref{qarad}) for $\rbh$, one finds that $\tbh \geq 0$ when $(X- X_{T-} ) ( X_{T+} -X  )\geq 0  $,  where
\be\label{xt}
X_{T\pm} = -X_{L\pm} + \sqrt{X_{L\pm}^2 + \at^2l^2}
\ee
Since $X_{L-} < X_{T-}$ and $ X_{L+} <  X_{T+}$,
requiring both $L^2 \geq 0$ and positive temperatures gives the allowed range of $X$,

\be\label{bigconstraintonx}
R_p \  : \quad X_{T-} \  \leq  X  \leq   X_{L+}
\ee
To have any values in this range it is necessary that 
\be\label{xlbound}
{\at l \over \sqrt{3} } \leq  X_{L-}
\ee
What is the physics of the range on $X$?  At the lower end, the black hole temperature vanishes but $L$ does not. Hence this is an extremal black hole
where the inner and event horizons coincide, but the cosmological and black hole event horizons are distinct. At the upper end, $L^2$ is zero,
so the black hole and deSitter horizons merge with the temperature of each going to zero.

To summarize, the physical parameter space for KNdS black holes is given by $X$ in the interval $R_p$ and $(a/l , q/l)$ in ${\cal P}$,
specified in equations  (\ref{bigconstraintonx}) and  (\ref{pset}) respectively.
 Note that the $3$-dimensional  parameter space ${\cal P}\times R_p$ is scale-invariant with respect to $l$,
consistent with the fact that the $l$ dependence of the metric simply appears as a conformal factor $l^2$, if the coordinates and parameters
are rescaled in units of $l$. 

The horizon radii, volume, and mass are then functions on this space given in  (\ref{qarad}), (\ref{volagain}), and (\ref{rotmass}).
It is helpful to look at Figure 2, which shows that in the allowed range of $X$ the black hole area and the mass  increase monotonically
with $X$ while the volume decreases. So as we discuss subsequent relations one can think of 
larger $X$ as larger area black holes and {\it vice-versa}.

\subsection{Entropy options and the structure of the parameter space}

\begin{figure}[t]
\begin{minipage}{0.5\textwidth}
        \label{features}
        \includegraphics[width=\textwidth]{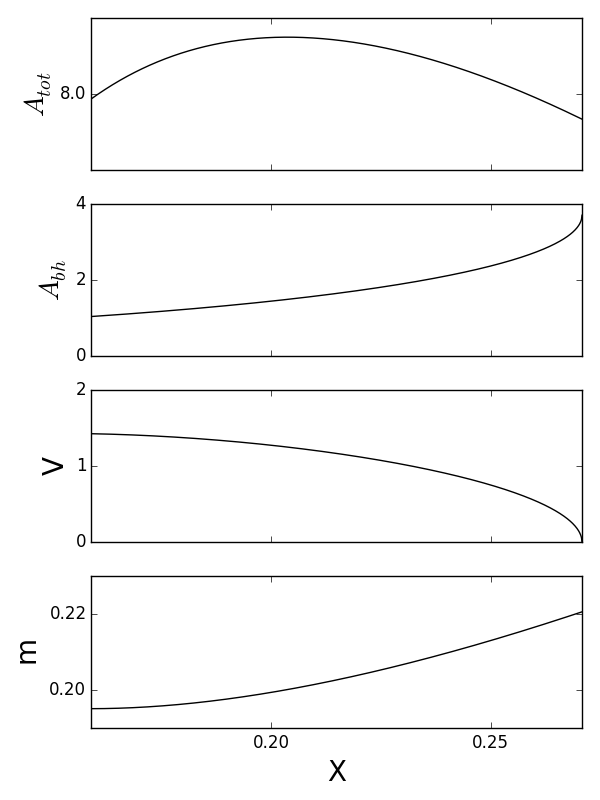}
        \caption{Total area, black hole horizon area, thermodynamic volume, and mass plotted as a function of $X$. The parameters $a, q$ are in 
        $\partial (\cal{SD} )$. Any horizontal line $A_{tot} = constant$ that intersects a small black hole also intersects a large one. } 
\end{minipage}
\begin{minipage}{0.6\textwidth}
        \label{parameter_graphic}
        \includegraphics[width=\textwidth]{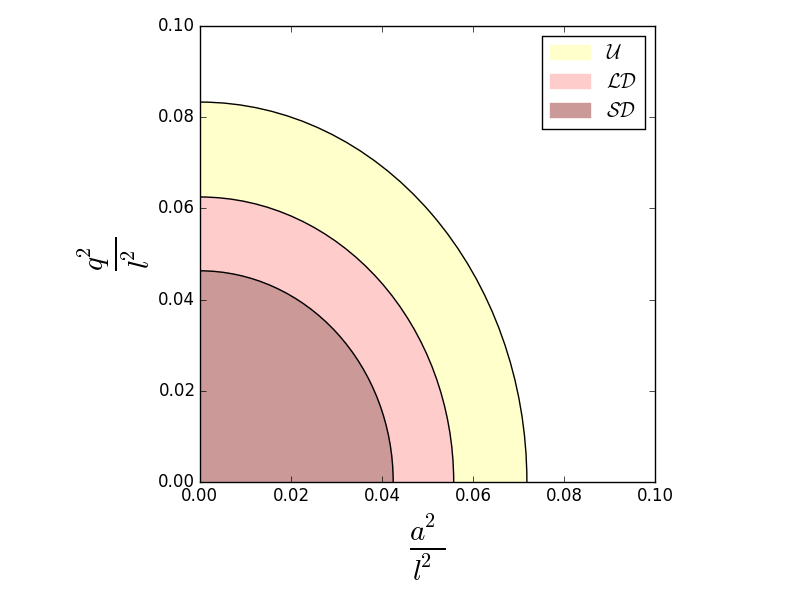}
        \caption{The physical parameter space ${\cal P}$ for $(a / l , q/ l )$. At each point there is a line segment $R_p$ (not shown)
        for the allowed values of $X$. In the region ${\cal U}$ each value of $S$ corresponds to a unique $X$; in ${\cal LD}$ (${\cal SD}$) large (small) black
        holes form degenerate $S$ pairs while $S$ is unique for small (large) black holes.  } 
        \end{minipage}
\end{figure}

\begin{figure*}
\centering
\begin{tabular}{cc}
\includegraphics[width=0.47\textwidth,height=0.27\textheight]{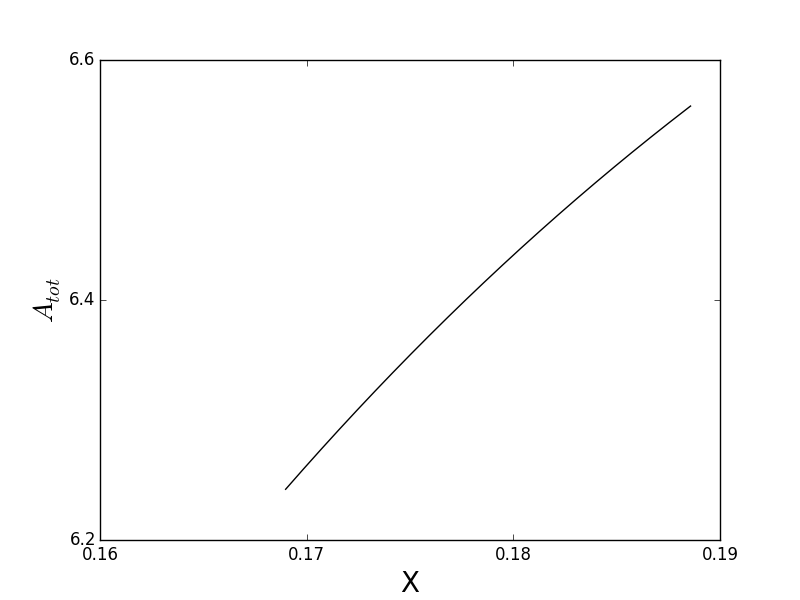} &
%\rotatebox{-90}{
\includegraphics[width=0.47\textwidth,height=0.27\textheight]{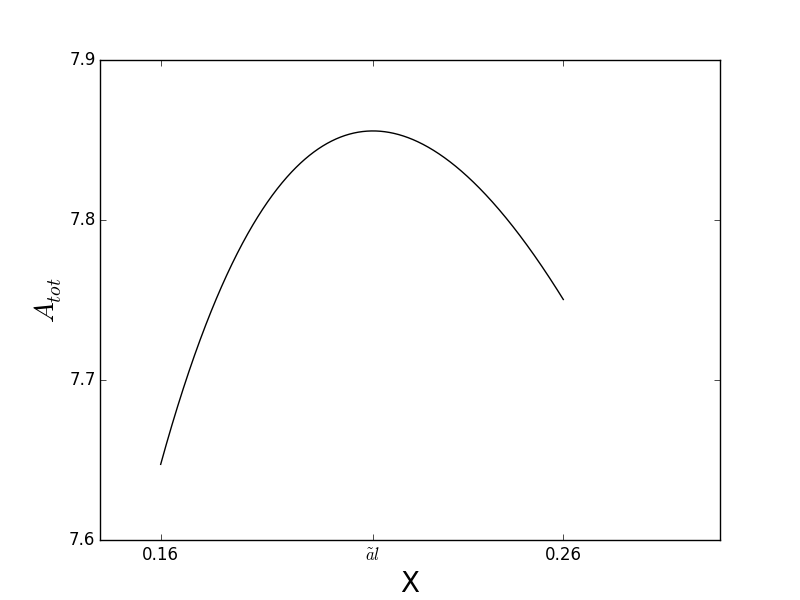}\\
\end{tabular}
\caption{{  $A_{tot} \ vs.\ X$:}
{\bf Non-degenerate:} On the left hand side $p=(a/l, q/l)$ is chosen from $\cal{U}$, see Figure 3. $A_{tot}$ does not reach its maximum, so 
        each $S$ corresponds to a unique black hole. {\bf Large-degenerate:} On the right hand side $p$ is chosen from $\cal{LD}$.
All large black holes are degenerate, $i.e.$,  are dual to a small black hole with the same $S$.
 }
\label{fig2}
\end{figure*}

%\begin{figure}[t]
%\begin{minipage}{0.5\textwidth}
%        \label{nondegenerate_area}
%        \includegraphics[width=\textwidth]{paper_plots/area_nondegenerate}
%        \caption{$A_{tot} \ vs.\ X$ with $a, q \in \cal{U}$. $A_{tot}$ does not reach its maximum, so 
%        each $S$ corresponds to a unique black hole.}
%\end{minipage}
%\begin{minipage}{0.5\textwidth}
%        \label{largebh_degenerate_area}
%        \includegraphics[width=\textwidth]{paper_plots/area_left_nondegenerate}
%        \caption{$A_{tot} \ vs.\  X$ with $a, q \in \cal{LD}$.  All large black holes are degenerate, $i.e.$,  are dual to a small black hole with the same $S$.}
%\end{minipage}
%\end{figure}
%

\begin{figure*}
\centering
\begin{tabular}{cc}
{\includegraphics[width=0.47\textwidth,height=0.27\textheight]{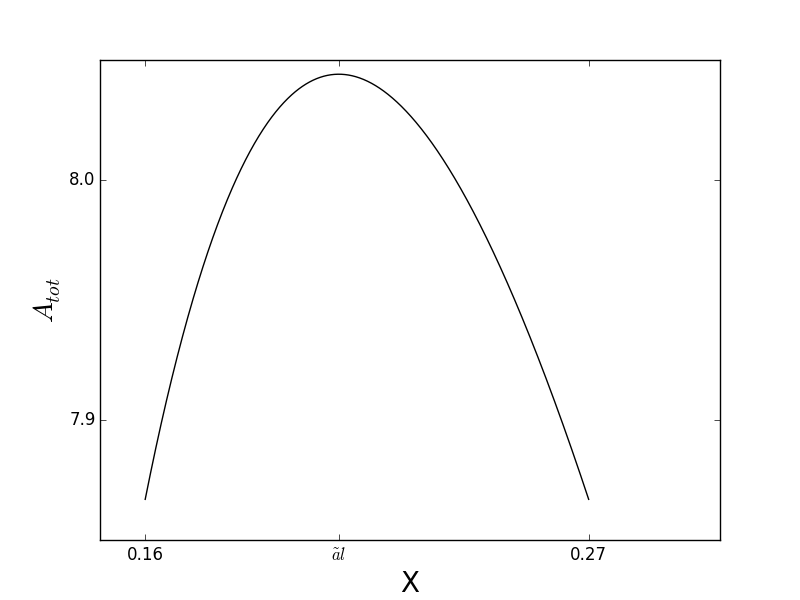}} &
\includegraphics[width=0.47\textwidth,height=0.27\textheight]{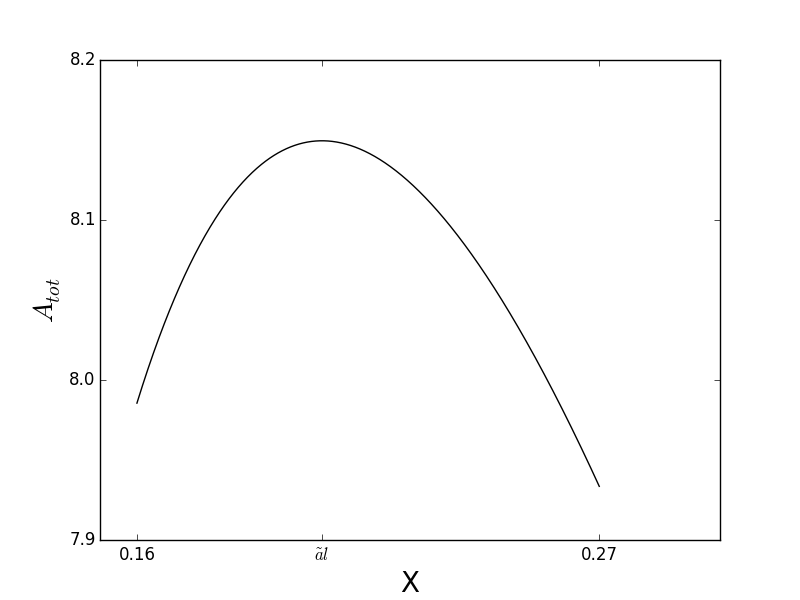}\\
\end{tabular}
\caption{{\bf  $A_{tot} \ vs.\ X$:}
{\bf Totally degenerate:} On the left hand side $p=(a/l, q/l)$ is chosen from $\partial (\cal{LS})$, see Figure 3. All black holes are dual to
another black hole with the same $S$.
 {\bf Small-degenerate:} On the right hand side $p$ is chosen from $\cal{SD}$.
All small black holes are degenerate, $i.e.$,  are dual to a large black hole with the same $S$.
 }
\label{fig3}
\end{figure*}

In this section we study properties of the total entropy $S={1\over 4} A_{tot}$. We will see that the parameter space ${\cal P}$ can
be divided into regions according to qualitative features of the total entropy  $S$. The results are summarized 
 in Figure 3, and we turn to the derivation of this diagram.
 
  One can read off of equation (\ref{totalarea}) that the entropy has the intriguing property that it is invariant under the map
\begin{equation}\label{map}
X \rightarrow X^\prime \equiv {\at^2 l^2 \over X}
\end{equation}
The maximum of $A_{tot}$ occurs at $X_{crit} =\at l$, which is the fixed point of this map. Hence
  for values of $A_{tot}$ less than the maximum,  two distinct values of $X$ will have the same total entropy. However, whether, or
 not, these values of $X$ correspond to physical black hole spacetimes depends on whether they are both in the allowed range  for $X$
 specified in equation (\ref{bigconstraintonx}), at the values of $(a/l , q/l) $ under consideration. 
 Hence our next step is to figure out the structure imposed on the 
  parameter space ${\cal P} \times R_p $ by the behavior of $S$. Specifically, let $ S_p $ be the set of values of the entropy
  that are obtained by  KNdS black holes at $p$. The map 
  \be\label{stox}
 \psi_p :  S_p \rightarrow X_p
 \ee
is either one-to-one or one-to-two. If $\psi$ is {\it 1-1} at $p$ then for each value of the entropy in $S_p$ there is a unique black hole. When $\psi_p$ is {\it 1-2}, then there
are two black hole spacetimes that have that have the same value $S$, that is, there are degenerate entropy states. We will refer to this behavior
as unique black holes ${\cal U}$, large black holes are degenerate ${\cal LD}$, or small black holes are degenerate ${\cal SD}$, according to whether
 $\psi_p$ is {\it 1-1} on the entire
range of $X$, {\it 1-2} for large $X$ and hence for large black holes, or {\it 1-2} for small $X$ and hence for small black holes. 
These different behaviors of $A_{tot}$ as a function of $X$,  depending on the value of $(a/l , q/l )$,  are illustrated in Figures 4 and 5.

The regions
in ${\cal P}$ divide nicely into rings, with the boundaries of the rings corresponding to special configurations, which are determined as follows.
Using $X_{L+}X_{L-} = \at^2 l^2 / 3$ and equation (\ref{xt}) gives
 $X_{T-}$ in terms of $X_{L+}$, and implies that $X_{T-} \leq \at l$.
 Since the maximum of $S$ occurs
at $X=\at l$,  there will be degeneracy at $p$ for some $X$ if $  \at l \leq X_{L+}$ at $p$. 

The smallest possible range of $X$ is the single value $ X =X_{T-}=X_{L+} = \at l / \sqrt{3} $.
 Using the expressions for  $X_{T-}$ and $X_{L+}$
 in equations (\ref{xl}) and (\ref{xt}) one finds that $(a/l , q/l)$ must satisfy $(1- a^2 / l^2  )^2 - 12 (\at / l )^2  = 0$
 which is just the boundary of the total parameter space ${\cal P}$. Hence all the KNdS spacetimes on $\partial {\cal P}$ have the same value of $X$.
  By continuity one expects that for some range of
$X_{L+} $ near $\at l / \sqrt{3} $ the entropy will likewise correspond to a unique black hole, so we want to find the inner boundary of the set 
${\cal U}$ within ${\cal P}$ that consists of black holes which are uniquely specified by $S$ and $p$. By the remarks in the previous paragraph, the inner boundary
of ${\cal U}$ is determined by the condition  $\at l = X_{L+}$, which translates into the curve
\be\label{bdryu}
\partial {\cal U}\   :\quad  16 {q^2\over l^2} = \left({a^2 \over l^2} - ( \sqrt{5} +2 )^2 \right)  \left({a^2 \over l^2} - ( \sqrt{5} -2 )^2 \right)
 \ee
Points $p$ inside this curve  will have a range of $X$ values that include both unique and degenerate entropy states. 
 Moving just inside $\partial {\cal U} $ one finds that the degenerate states occur at the larger values of $X$, which are larger area black holes and are
 denoted ${\cal LD}$. The region at the center of ${\cal P}$ contains the points at which the degenerate states occur at small $X$, which are smaller area black holes,
 or ${\cal SD}$. The boundary curve between the large and small degenerate states is when all $X$ values have isentropic partners. This occurs when
 the endpoints are dual to each other satisfying $X_{L+}^\prime = X_{T-}$. One finds that this condition becomes
 $X_{L+} = \sqrt{{5} \over3}\at l$ and the curve in ${\cal P}$ on which all states are degenerate is
 \begin{equation}\label{bdrys}
\partial ({\cal LS}) \  : \quad
108{q^2 \over l^2 } + 118 {a^2 \over l^2 }  - 5{a^4 \over l^4 }  - 5 =0
\end{equation}
 Collecting these results, we have the following division of the parameter space ${\cal P}$: 

 $\underline{   \partial {\cal P} :}  \  \  X_{T-}=X_{L+} =\at l / \sqrt{3}  $\bk
  In this case the range of $X$ has shrunk to one value, indicating that
  the three horizons coincide and the temperatures are zero. 
  This configuration was called the ``ultra extreme black hole" in the analysis of coinciding horizons of charged black holes in deSitter \cite{Brill:1993tw}. 
The algebraic equation  for the perimeter is given by the boundary of (\ref{pset}). 
  
  $\underline{  {\cal U} : \  Unique\  states: }\quad \at l / \sqrt{3} < \  X_{T-},  \    X_{T+} < \  \at l$.\bk
  The upper limit for $X$ is less than where the area curve reaches a maximum.
  All black holes have a unique value of $S$. ($\psi$ is {\it 1-1} on $R_p$.) The inner boundary is given by the curve (\ref{bdryu}), See Figure 4.
  
   $\underline{ {\cal LD} : \  Large \  Degenerate:} \quad  \at l / \sqrt{3} < \  X_{T-} ,\  \at l \   < X_{T+} < \   \sqrt{5\over 3}  \at l$.\bk
     The upper limit for $X$ has increased to include the value where $S$ is maximum. 
   All large black holes are dual to a smaller black hole with the same $S$. ($\psi$ is {\it 1-1} at small $X$ and is {\it 2-1} at large $X$ on $R_p$.) 
   See Figure 4.
   
     $\underline{ \partial  { (\cal  LS}) : \ All\  Degenerate:} \quad   X_{T-} = \sqrt{3\over 5}  \at l ,\  X_{L+} =\   \sqrt{5\over 3}  \at l$.\bk
     At the boundary between ${\cal  LD}$, where large black holes are degenerate but not all the small ones are, and ${\cal  SD}$,
     where small black holes are degenerate but not all the large ones are, is the balanced situation when
     every black hole is dual to another black hole. ($\psi$ is {\it 1-2} on $R_p$.) The equation for this curve is given in (\ref{bdrys}).
     See Figure 5.
   
    $\underline{ {\cal SD} : \  Small \  Degenerate:} \quad   \sqrt{3\over 5}  \at l   < X_{T-} < \at l   , \   \sqrt{5\over 3}\at l \ < X_{L+}   $.\bk
    The upper and lower limits for $X$ have increased so that some large black holes no longer have dual states.
     All small black holes are dual to a larger black hole with the same $S$. ($\psi$ is {\it 1-2} at small $X$ and is {\it 1-1} at large $X$ on $R_p$.) 
     See Figure 5.

The structure of ${\cal P}$ is summarized in Figure 3, which we see splits neatly into
rings that delineate 
  whether each value of the entropy corresponds to a unique KNdS black hole, or if some values of $S$
  correspond to two black holes. 
The graphs in Figure 2 illustrate that the  mass and the area of the black hole horizon are  monotonically related to area product $X$.
For $a=q=0$ it  simple to show that $m$ and $\abh$ increase monotonically with $X$ and with each other.
 A more tedious, but straightforward, calculation shows that when $a$ and $q$ are nonzero it is still the case that  $\abh$ and $m$ increase 
monotonically with $X$, in the physical parameter space. 
Hence black holes with large areas do indeed correspond to 
black holes with large mass, and when
 reading  plots with independent variable $X$ one can think of increasing $X$ as  increasing black hole area and mass.

In classical thermodynamics, a system evolves to maximize its entropy, and in equilibrium, the components of a system  have the same temperature. 
The thermodynamic properties of  charged deSitter black holes with no rotation are consistent with our non-gravitating thermodynamic expectations, as
 the equal temperature black holes are also maximum entropy black holes. These occur when 
 $X= ql$ and satisfy $m=q$.  However, when
rotation is added these these two properties no longer coincide. The maximum of the entropy occurs at  $X_{maxS}=\at l$ which does
not equal $X_{eq}$ given in equation (\ref{xeq}) unless $a=0$.  At the  maximum entropy  black hole the mass is given by
\be\label{maxsmass}
m_{crit}  = \sqrt{ a^2 + q^2 } \sqrt{ 1-  a^2 / l^2} 
\ee 
where we have explicitly written out the charge dependence to highlight how $m=q$ generalizes when $a$ is non-zero. The first factor
is the ``norm" of the conserved charges, and the second factor has the form of a gravitational redshift. So when
considering what physics of the KNdS metrics persist in a situation like inflation when $\Lambda$ is actually dynamical, it may
be that maximizing $S$, or evolving to equal temperatures, or some compromise between the two, is the key principle.

\section{Embedding diagrams: picturing isentropic black holes}\label{pictures}

\begin{figure*}
\centering
\begin{tabular}{cc}
{\includegraphics[width=0.47\textwidth,height=0.27\textheight]{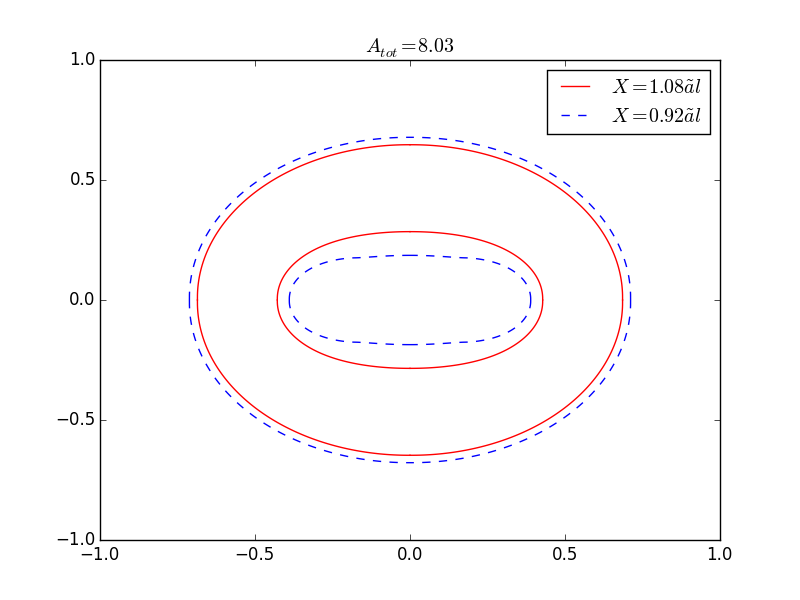}} &
%\rotatebox{-90}{
\includegraphics[width=0.47\textwidth,height=0.27\textheight]{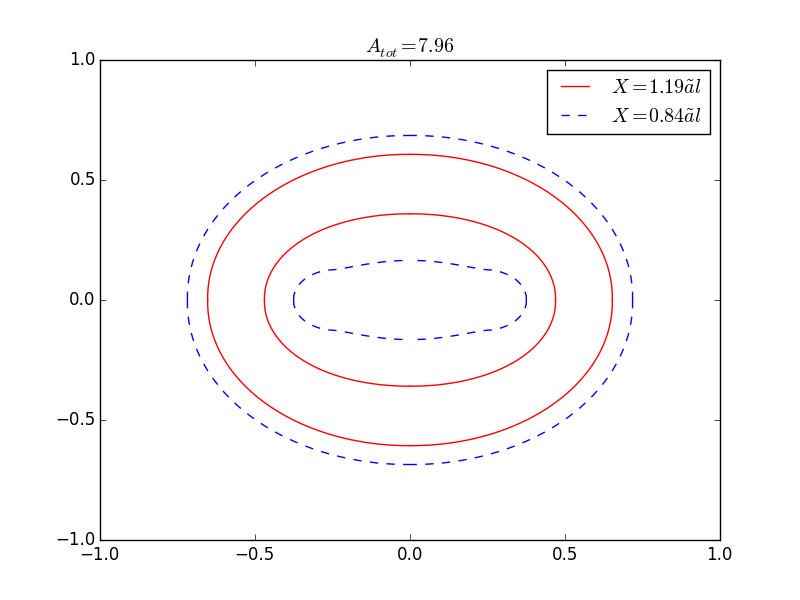}\\
\end{tabular}
\caption{{\bf  Embedding diagrams of isentropic horizons:}
The black hole (smaller) and cosmological (larger) horizons are drawn for a pair of isentropic spacetimes, dashed for the larger $V$ spacetime and
solid for the smaller $V$ dual partner.
 On the left hand side both systems are located near the fixed point $X=\tilde{a}l$, and are thus similar in configuration.
 On the right hand side $X $ chosen further from the fixed point so that the
         two systems exhibit differences in size and configuration. The smaller black hole reveals a plateau. The parameters are chosen from
         $ \partial (\cal{LS} )$, where all black holes are degenerate.
          }
 \label{fig4}
\end{figure*}

The isentropic KNdS black hole spacetimes have the same total horizon areas but different values of $X$, so it is 
interesting to describe further how they differ geometrically, and to see what dual pairs look like. Choose values of $(l,a,q)$ and a
dual pair in the physical parameter space, as illustrated  in Figure 6. Let
 $X_*  < \at l$, and let's compare the dual spacetimes  having $X_*$ and $X_*^\prime$ respectively.
 $X_*$ is the smaller area black hole of the pair. It is simple to 
 check that $m_* < m_*^\prime$. With a bit more algebra one also finds that
 $V_*  > V_*^\prime$, as long as  $X_*$ and $X_*^\prime$ are in the allowed parameter space. Hence the dual
 rotating black holes  have the same cosmological constant, charge, spin, and  total horizon
areas, but different thermodynamic volumes between the horizons and different masses. The smaller area black hole spacetime $X_*$ has the 
smaller mass and larger volume between the horizons, so heuristically this is the ``lower density" state.

To see what the dual pairs look like
we construct embedding diagrams. 
Each diagram is constructed so that
the metric of the embedded two-dimensional surface in flat three dimensional space matches 
the KNdS metric (\ref{rotmetric}) restricted to the two-dimensional surface $t=constant$ and $r=r_h$ . Because of the cylindrical symmetry one can use standard
 cylindrical coordinates $(z, \rho ,\phi )$ in the flat space. We then suppress  the symmetry coordinate $\phi$.  The coordinate
 $\theta$ in the KNdS metric can be used as the parameter along the embedded curve. Identifying $\rho = \sqrt{g_{\phi\phi} }$, the equation for the surface is
 \be\label{emsurf}
  (  {dz \over d\theta } )^2  =  g_{\theta \theta } - ( {d\over d\theta } \sqrt{\rho } )^2 
 \ee

For a given spacetime (fixed $l,a, q, X$) we plot the black hole and the cosmological horizons. Each embedding diagram in Figure 6 contains 
the diagrams for an isentropic pair of KNdS spacetimes, one drawn with dashed lines and the dual one drawn with solid lines.
The sum of the areas of the dashed horizons is equal to the sum of the areas of the solid horizons. 
In left hand side of Figure 6, the entropy is chosen to be near the maximum value, that is, $X$
 and $X^\prime = \at^2 l^2 / X$  close to the fixed point $X_{Smax} = \at l$.
Hence each horizon of the unprimed spacetime is quite similar to the dual horizon in the primed spacetime. The right hand side of Figure 6 provides a contrast
where the value of the total entropy is different enough from the maximum that one can see qualitative differences
between the horizons of the dual spacetimes. The smaller black hole horizon shows the development of a plateau type feature,
giving it a squashed four-leaf clover shape. We have found that this is a consistent characteristic of the near-extremal black holes as
$\rbh$ approaches $r_{in}$. We have checked that $\Lambda =0$
Kerr black holes look similar in the near extremal limit, so the plateau at intermediate $\theta$ is not due to $\Lambda$.
This effect can be checked analytically by examining the right hand side of (\ref{emsurf}) as a function of $a/ \rbh$.
The extremal black hole is $a/ \rbh \rightarrow 1$. One finds that if $a/ \rbh$ is small then $dz/d\theta$ only goes to zero at the pole
$\theta =0$, but if $a/ \rbh$ is close to one, $dz/d\theta$ does get small at an angle intermediate between the equator and the pole.
 
 We also provide animations of sequences of embedding diagrams of the horizons for the KNdS \cite{video}
spacetimes, in which $A_{tot}$ is used as a time variable.
The video starts with a sequence for a point in ${\cal U}$ so that each value of $A_{tot}$ defines a unique black hole. Second is
 a sequence for a point in ${\cal LD}$ so that large black holes have a isentropic partner, and third is a sequence 
 from $\partial ( {\cal LS} )$. Once the degenerate area  regime
 is reached the dual pairs are displayed together, with the sequence ending at the maximal $A_{tot}$. Lastly is an example
 from the region ${\cal SD}$ in which dual spacetimes occur for the small black holes.

\section{Compressibility and isentropic phase transitions}\label{compress}

We have seen that for a given value of the cosmological constant, KNdS black holes may either be uniquely specified by the 
total entropy, or there may be two distinct black holes with the same $S_{tot}$, depending on the values of $(a , q)$.
When there are degenerate states  one branch has  larger thermodynamic volume and smaller mass,  and the other branch has smaller volume and larger mass.
In this section we show that the compressibility at constant total entropy $S$ is a diagnostic that distinguishes the two phases, and diverges at the critical point
where the two phases coincide, which is the maximal entropy configuration for the given values $l, a,$ and  $q$. Reference
\cite{Dolan:2013dga} computed the compressibility for rotating AdS black holes. 

In a standard thermodynamic system, the compressibility at constant entropy is defined
as $\beta _S = - {1\over V} \partial V /\partial P )_S $. 
For  familiar materials with positive
 pressure, as the pressure increases  the volume decreases, and  so $\beta_S$ is positive. In the KNdS spacetimes the pressure is provided by the cosmological
constant  $P =-3/(8\pi l^2)$, and $\beta_S$ becomes
\begin{equation}\label{comprdef}
\beta _S = -{4\pi \over 3} { l^3 \over V}  \left( {\partial V \over \partial l} \right)_S
\end{equation}
When $q=a=0$ the calculation is particularly transparent.  Equation
(\ref{horarea}) can be rewritten as  $(3V /4 \pi ) ^2 = \ 3l^4 S_{tot} /\pi - 2l^6$.
Taking the derivative with respect to $l$ at fixed $S_{tot}$ gives $V (\partial V / \partial l)
=- 6 l^3 (l^2 -S_{tot}) /\pi $. The results of Section (\ref{SdSbhs}) show that $S_{tot}$ is always less than its pure deSitter value of $ l^2$.
 So even though the pressure provided by the cosmological 
constant  is negative, the compressibility is positive and given by
\begin{equation}\label{compr}
\beta_S = {8\pi l^2 \over 3 \mu ^ 2} ( 1-\mu ^2 )
\end{equation}
where $\mu = 3V/(4\pi l^3 )$. As the black hole and cosmological horizons approach each other, the volume goes to zero
and the compressibility diverges. 

 In the general case we use
equation (\ref{volagain}) to compute $\partial V / \partial l $ at fixed $S$, $a$, and $q$. There is both explicit and implicit dependence on
$l$, that is,
\begin{eqnarray}\label{comprgen}
\beta _S  & = & - {4\pi l^3 \over 3}  \left\{ -  \left[ {1\over L^2} \left(3-{\at ^2 l^2 \over X^2 } \right) + {\at^2 \over  X} {1\over  X- \at ^2  } \right]  \left( {\partial X\over \partial l} \right)_{S} 
\right.  \\  \nonumber
 &+&   \left.  {2(1+ 2 a^2 / l^2 )\over (1+  a^2 /  l^2 )}  \  + {l \over L^2 } \left(1- {\at^2 \over X} \right)\right\}  \\ \nonumber
\end{eqnarray}
The derivative  $\left( \partial X / \partial l \right)_S $ is determined by requiring $\delta S=0$. Using equation (\ref{totalarea}) one finds
\be\label{deriv}
\left( {\partial X\over \partial l} \right)_S = {2l \over \gamma }{X^2\over ( X^2  - \at^2 l^2   ) } 
\left(1-{a^4 \over l^4} +{a^2 \over l^4 } X - {\at^2 \over X} \right)
\ee
Expression (\ref{comprgen}) with (\ref{deriv})  reduces to (\ref{compr}) when $a=q=0$.
 Note that $( {\partial X\over \partial l} )_S$ diverges as $X$ approaches the critical value $X_{crit} = \at l$, which is where the entropy is maximum.
The other terms in (\ref{comprgen}) remain finite in this limit, and we find 
\be\label{compr}
\beta _S \sim {4\pi \over 3} { l^3 \at \over X-  \at l } 
\ee
where we have used $a/l , \at / l \ll 1$ to get a simple expression. Let $V_{crit }$
be the value of $V$ at $X_{crit}= \at l$. Then near the critical point $\beta_S$ 
can be rewritten in terms of the dimensionless volume fluctuation $v-v_{crit} = {3\over 4\pi } ( V - V_{crit} )/l^3 $.  One finds
\be\label{comprv}
\beta _S \sim  {8\pi \over 3} {l \at \over v_{crit}-v } \  
\ee
Hence $\beta_S$ changes sign at $v_{crit}$, where it diverges. Larger area black holes, which have $X>\at l$ and $v< v_{crit}$ have
positive compressibility, but the situation is reversed for smaller area black holes.
Graphing  $\beta_S$ for the entire range of $X$ shows that it is negative for $X_{T-}< X<\at l$ and is positive for $\at l < X < X_{L+} $,
though we have not shown this analytically. This result says that when the horizons are close together and so $v$ is small, if $\Lambda$ increases then in
order to keep the total area constant the volume
must increase, that is, the horizons get further apart.  On the other hand when  when the black hole is small and so $v$ is large, if $\Lambda$ increases
the volume must decrease, which means the horizons move closer together. In either case the horizons move towards an intermediate
volume configuration upon an increase in $\Lambda$. Of course the directions are reversed if $\Lambda$ decreases.

To summarize, we have seen that
in the ${\cal LD}$ and ${\cal SD}$ regions of parameter space, at each value of $(a/l , q/l )$ the family of KNdS black holes
includes a  maximal entropy black hole which divides the family into a branch with smaller $\abh$, smaller $m$ and larger $V$,
and those with larger $\abh$, larger $m$, and smaller $V$. The two branches merge at the maximal entropy state where the compressibility
diverges. For a range of the states close to the maximum, each state on one branch is dual to a state on the other branch
with the same $S$. This picture is suggestive of a phase transition between large and small black holes, or more heuristically, high and low ``density"
phases with density $\sim m/V$. Near the critical point, the properties of the  two phases become more similar, and
isentropic fluctuations would mix the two phases.
Of course in this current work we have only included variations amongst the KNdS spacetimes. It would be interesting to study more general fluctuations
about a maximal entropy black hole and see if it continues to have properties of a critical point in phase space.

\section{Discussion}\label{end}

In this paper we have introduced a new parameterization for KNdS black holes. We have shown that there are practical benefits,
specifically the thermodynamic quantities are given as functions of the cosmological constant, the spin, the charge, and the new 
area product parameter. As a result the cosmography is more accessible, and we have
illustrated interesting qualitatively different behaviors of the temperatures and entropies. Though several examples have
been given there is clearly room for more exploration of the geometry of KNdS black holes. 

Our results lead to  several further questions.
 The symmetry  of the entropy  under inversion of $X$ significantly simplifies the analysis of the black hole  properties, 
 and it is intriguing that the maximal entropy state is the fixed point of this map, but
  we have not yet found a more fundamental underlying principle for this symmetry. It may be that the  dS/CFT point(s) of view
 will shed light on this issue.  It would be interesting to probe  the
 maximal entropy spacetime by analyzing fluctuations about it, and understand if this does indeed have the properties
 of a critical point. A related, but 
 challenging, issue is to study the role of quantum mechanical black holes and
 cosmological particle production in the KNdS metrics, extending the analysis in \cite{Kastor:1993mj}.
  We have seen that the maximal
 entropy state does not coincide with the equal temperature state, so which one ``wins" in a semi-classical evolution? 

Another set of questions has to do with instanton calculations.
 In the $\Lambda =0$ case  \cite{Garfinkle:1990qj,Dowker:1993bt} 
it was found that the production rate is controlled by the  factor $\exp (-S_{bh})$. 
 Calculations of pair production of KNdS black holes  \cite{Mellor:1989wc,Mann:1995vb,Booth:1998gf,Booth:1998pb,Dias:2003st,Dias:2004rz}
 and nucleation of dS bubbles around black holes  \cite{Gregory:2013hja,Burda:2015isa,Burda:2015yfa}
  have found that this result generalizes to $\exp [ -( S_{bh} + S_{cos}) ]$. Hence the degeneracy 
  of KNdS black holes may imply an enhancement of the rates for certain processes. In particular, near the maximum entropy state
  there are a continuous set of black holes with almost the same entropy and almost the same parameters. This geometry may
  lead to bump in the rate for producing black holes with the same $l, a$ and $q$ but a spread in masses.
   
  One of the motivations for studying KNdS black holes is to understand the role of black holes in the early universe. In the inflationary universe 
 the cosmological constant is in fact dynamical, and incorporating slow changes in $\Lambda$ should be tractable in a perturbative approach,
 as has been carried out for SdS black holes with a slowly changing scalar field \cite{Chadburn:2013mta}. In addition, 
  in many inflationary models the accelerated expansion is a power law rather than the
 true deSitter exponential.  Do the qualitative features of the behavior of the thermodynamic quantities for KNdS to continue to hold? 
 One might expect that the critical property is that the spacetime has a cosmological horizon,
 so that the causal structure of the observable universe is the same as in deSitter. Positive evidence along these lines is provided by
studies of the global stability of accelerating spacetimes
\cite{Friedrich:1986} \cite{Andersson:2004} \cite{Ringstrom:2008} \cite{Ringstrom:2009} \cite{Luo:2012bn}, which show that the
 localization of the future development of data on a patch of an initial value slice is a crucial element
 in analyzing stability, whether  the acceleration is exponential or power law. 
 
Lastly, we note that another issue is to understand the physics of the sign change of
 $\beta_S$. In classical thermodynamics,
fluids that have positive compressibility behave as expected. A fluid with negative compressibility, however, expands when compressed and contracts when tension is applied.
Recently, changes in the sign of the compressibility have been seen in metamaterials like foams and in certain acoustic metamaterials \cite{Fang,Ding}.
In the latter case a change in the frequency of incident sound waves can produce a fluid with negative compressibility. One can speculate that 
if research on the dS/CFT correspondence evolves along the lines of the AdS/CFT correspondence that KNdS black holes could provide a gravitating
dual for such materials - or not!

\section{Acknowledgements} The work of JM and GS was supported in part by the Faculty First Year Seminar Program  and the
 Commonwealth College of the University of Massachusetts. JT thanks David Kastor, David Kubiznak, Narayan Menon,
and Shubha Tewari for useful conversations.

%\begin{thebibliography}{99}
\providecommand{\href}[2]{#2}\begingroup\raggedright\endgroup

\end{document}